# An unusual continuous paramagnetic-limited superconducting phase transition in 2D NbSe$_2$

Egon Sohn[1,2], Xiaoxiang Xi[1,3], Wen-Yu He[4], Shengwei Jiang[1,2], Zefang Wang[1,2], Kaifei Kang[1,2], Ju-Hyun Park[5], Helmuth Berger[6], László Forró[6], Kam Tuen Law[4], Jie Shan[1,2,7]*, Kin Fai Mak[1,2,7]*

[1]Department of Physics, The Pennsylvania State University, University Park, Pennsylvania 16802, USA
[2]Department of Physics and School of Applied and Engineering Physics, Cornell University, Ithaca, New York 14853, USA
[3]National Laboratory of Solid State Microstructures, School of Physics, and Collaborative Innovation Center of Advanced Microstructures, Nanjing University, Nanjing 210093, China
[4]Department of Physics, Hong Kong University of Science and Technology, Clear Water Bay, Hong Kong, China
[5]National High Magnetic Field Laboratory, Florida State University, Tallahassee, Florida 32310, USA
[6]Institute of Condensed Matter Physics, Ecole Polytechnique Fédérale de Lausanne, 1015 Lausanne, Switzerland
[7]Kavli Institute at Cornell for Nanoscale Science, Ithaca, New York 14853, USA

*Correspondence to: jie.shan@cornell.edu, kinfai.mak@cornell.edu

**Time reversal and spatial inversion are two key symmetries for conventional Bardeen-Cooper-Schrieffer (BCS) superconductivity [1]. Breaking inversion symmetry can lead to mixed-parity Cooper pairing and unconventional superconducting properties [1-5]. Two-dimensional (2D) NbSe$_2$ has emerged as a new non-centrosymmetric superconductor with the unique out-of-plane or Ising spin-orbit coupling (SOC) [6-9]. Here, we report the observation of an unusual continuous paramagnetic-limited superconductor-normal metal transition in 2D NbSe$_2$. Using tunneling spectroscopy under high in-plane magnetic fields, we observe a continuous closing of the superconducting gap at the upper critical field at low temperatures, in stark contrast to the abrupt first-order transition observed in BCS thin film superconductors [10-12]. The paramagnetic-limited continuous transition arises from a large spin susceptibility of the superconducting phase due to the Ising SOC. The result is further supported by self-consistent mean-field calculations based on the *ab initio* band structure of 2D NbSe$_2$. Our findings establish 2D NbSe$_2$ as a promising platform for exploring novel spin-dependent superconducting phenomena and device concepts [1], such as equal-spin Andreev reflection [13] and topological superconductivity [14-16].**

In conventional BCS superconductors, Cooper pairs are even-parity singlet [17], which yields nearly zero spin susceptibility at low temperatures. The paramagnetic-limited superconductor-normal metal transition is thus an abrupt first-order transition at the upper critical field [10] (Fig. 1a). This has been experimentally verified, for instance, in very thin superconducting aluminum [10, 11] and beryllium [12] films under an in-plane magnetic field, for which orbital depairing is suppressed. In non-centrosymmetric superconductors, which lack a center of inversion in the crystal structure, the presence of antisymmetric spin-orbit coupling (SOC) can give rise to a wealth of novel properties, including the superconducting magneto-electric effect [1], spin Hall effect [1] and helical phases [1, 18]. In particular, the spin susceptibility of the superconducting phase can become significant compared to the normal-state value [19-21] and lead to a continuous paramagnetic-limited transition in the zero-temperature limit [22] (Fig. 1a). Such a transition, however, has not yet been observed in any systems. The recently emerged non-centrosymmetric superconductors 2D $NbSe_2$ [6-9] and gated $MoS_2$ [23, 24] possess unique Ising SOC, which pins the electron spins to the out-of-plane direction. They also have sample thickness much smaller than the bulk penetration depth, and therefore, a significantly suppressed orbital response to an in-plane magnetic field [7, 9, 23, 25]. In contrast to the well known non-centrosymmetric superconductors with Rashba SOC [1-5], in which orbital effect is often important, atomically thin superconductors with Ising SOC provide an ideal platform for the investigation of the paramagnetic-limited phase transition and other pure spin-dependent superconducting phenomena and device concepts.

Single-layer $NbSe_2$ consists of a layer of Nb atoms sandwiched between two layers of Se atoms in a trigonal prismatic structure [9, 26]. It has out-of-plane mirror symmetry and broken inversion symmetry. Electrons are subject to strong out-of-plane SOC fields (opposite at opposing crystal momenta), and are effectively Ising spins [9, 13, 23, 24]. The Ising spins are largely preserved in each monolayer of multilayer samples due to the weak interlayer interactions [9] (Fig. 1b). Recent experimental advances in high-quality atomically thin $NbSe_2$ samples have unveiled the coexistence of charge-density-wave order and superconductivity down to the monolayer limit [6, 8, 9, 26], in-plane upper critical fields far exceeding the spin paramagnetic limit of the BCS theory [9], and a Bose metal phase [7]. The earlier studies [9, 23, 24] based on resistance measurements, however, do not have direct access to the spectroscopic information and the superconducting order parameter [13, 15, 18]. Here we report the first tunneling measurement of a non-centrosymmetric superconductor in the paramagnetic-limited regime, and the observation of an unusual continuous paramagnetic-limited phase transition in 2D $NbSe_2$ under a high in-plane magnetic field. Our result provides strong evidence for a significant spin susceptibility of the superconducting phase in the zero temperature limit originated from the Ising SOC.

Figure 1c shows the optical image and energy diagram of a typical normal metal-insulator-superconductor (NIS) junction that has been employed in this experiment. It consists of a thin layer of platinum or gold electrode, a thin layer of tunnel barrier and a superconducting $NbSe_2$ layer. Two different tunnel barrier materials, aluminum oxide ($AlO_x$) and few-layer $MoS_2$, have been tested. Both types yielded similar results. Bilayer and trilayer instead of monolayer $NbSe_2$ have been studied because of their significantly higher chemical stabilities and sheet critical current densities. Because of the weak interlayer interactions, results for atomically thin samples are similar. Below we present the trilayer result and include the bilayer result in Supplementary Sect. 4 and 5. Both the four-point resistance $R$ and the differential tunneling conductance spectrum $G(V)$ (= d$I$/d$V$, where $I$ is the tunneling current and $V$ is the bias voltage) have been measured on the same devices. These quantities normalized by the normal-state values ($R_N$ and $G_N$) are shown in Fig. 1 and 2. Raw differential conductance spectra are included in Supplementary Sect. 4. Details on the device fabrication and electrical measurements are provided in Methods and Supplementary Sect. 1 - 3.

The basic characterization of 2D NbSe$_2$ superconductors by four-point resistance measurements is shown in Fig. 1e and 1f. The critical temperature determined from the onset of the transition (90% of the normal resistance) is $T_C \approx 5.8$ K (Fig. 1e, top panel). The in-plane critical field $H_{c2}^{\parallel}$ determined from the onset of the transition increases with decreasing temperature and saturates at ~ 38 T (Fig. 1f). This value is about 3.5 times the Pauli paramagnetic limit $H_P$ ($\approx 1.84\ T_C$) of the BCS theory [17], which supports the preserved Ising spins in trilayer NbSe$_2$ as a result of weak interlayer interactions [9]. Meanwhile, the transition width (shaded region of the fields between 90% and 10% of the normal-state resistance) increases with temperature due to the enhanced thermal fluctuations [10, 17]. It approaches about 2 T at low temperatures, likely limited by sample inhomogeneities. These results agree well with the reported studies [6, 7, 9, 26].

We now investigate the superconducting gap as a function of temperature and in-plane field by tunneling spectroscopy. The differential tunneling conductance spectra under zero magnetic fields are shown in Fig. 1d (symbols) for several representative temperatures. At low temperatures, they consist of two symmetric peaks that are stemmed from tunneling of normal electrons into the electron and hole branches of the quasiparticles in the superconductor [17] (Fig. 1c). The separation between the peaks corresponds to the superconducting gap $2\Delta \approx 2$ meV. Finite in-gap conductance is also observed and is caused by Andreev reflection at junctions with finite transparency [17, 27]. As temperature increases, the quasiparticle peaks broaden and their separation decreases, indicating the closing of the superconducting gap. For a more quantitative analysis of the tunneling spectra, we compare them to the Blonder-Tinkham-Klapwijk (BTK) model that includes both the single quasiparticle tunneling and the Andreev process in an NIS junction of arbitrary barrier strength [17, 27] (solid lines, Fig. 1d). The dimensionless barrier strength, gap energy and quasiparticle lifetime have been used as free fitting parameters. (For details refer to Supplementary Sect. 5.1 and 5.2.) The agreement between experiment and theory is generally very good except the small dips observed outside the superconducting gap unaccounted. These features likely arise from heating of local constrictions in the junction [27]. For typical junctions used in this experiment, the barrier strength (~ 0.5 – 1.5) is moderate. The temperature dependence of the gap is depicted in Fig. 1e (symbols) and can be described by the BCS theory [17] (solid line). The extracted $T_C$ (dashed vertical line) is consistent with the value from the resistance measurements. The extracted zero-temperature gap size $2\Delta_0 \approx 4.3 k_B T_C$ (where $k_B$ is the Boltzmann constant) is higher than the usual $3.52 k_B T_C$ for a BCS superconductor [17], likely due to strong-coupling corrections [28].

The behavior of the superconducting gap in an in-plane magnetic field is illustrated in Fig. 2. The contour plot in Fig. 2a is the differential conductance as a function of bias voltage $V$ (bottom axis) and in-plane field $H_\parallel$ up to about 40 T (left axis) at 0.3 K ($\approx 0.05 T_C$). Figure 2b shows the representative tunneling spectra (symbols) together with comparison to the BTK model (solid lines). The BTK analysis here has neglected both the orbital effect and the Zeeman effect (see Methods for justification based on the estimated orbital-limited upper critical field). The extracted gap as a function of $H_\parallel$ is summarized in Fig. 2c and the other parameters, in Supplementary Sect. 5.3, respectively. The figure also includes results at several elevated temperatures. (See Supplementary Sect. 5.5 for raw spectra and analysis.) As a comparison, we have included the classic example of superconducting aluminum thin films under an in-plane field [10, 11] in Fig. 2d (tunneling spectra) and Fig. 2e (superconducting gap extracted from the data in ref. 11 by our BTK analysis). Our results differ drastically from the behaviors of BCS thin film superconductors: (1) The two quasiparticle peaks have survived up to $H_\parallel > 30$ T, which far exceeds the Pauli paramagnetic limit of the BCS superconductors [17]. (2) The two quasiparticle peaks exhibit negligible Zeeman splitting even under very high fields unlike the significant Zeeman splitting observed in the BCS superconductors [10, 11]. (3) As $H_\parallel$ increases, the superconducting gap drops continuously to zero at the upper critical field in the zero-temperature limit, indicating a continuous phase transition, in stark contrast to the abrupt first-order phase transition observed in BCS superconductors [10-12].

The observed continuous paramagnetic-limited superconductor-normal metal transition in 2D NbSe$_2$ is rather unique. Table 1 summarizes all currently available experiments on the order of the paramagnetic-limited transition. They range from metal thin films [10-12] to quasi-2D heavy fermion [29, 30], organic [31] and iron pnictide [32] systems such that orbital depairing is strongly suppressed under a magnetic field in order to reach the paramagnetic-limited regime. In the low-temperature limit, they all exhibit a first-order transition in contrast to 2D NbSe$_2$. We note that non-centrosymmetric superconductors with Rashba SOC in principle can also exhibit a continuous paramagnetic-limited transition, but under an out-of-plane field (Supplementary Sect. 6.3). The orbital depairing under this configuration, however, makes it almost impossible to reach the paramagnetic-limited regime.

To understand our experimental observations, we consider the free energies of the normal and superconducting states in the zero-temperature limit when all real quasiparticle excitations are frozen. Including only the spin paramagnetic effect, the free energy can be expressed as $F_N = -\chi_{\parallel N} H_\parallel^2/2$ for the normal state, and $F_S = -N_0 \Delta_0^2/2 - \chi_{\parallel S} H_\parallel^2/2$ for the superconducting state (ref. [1, 17]). Here $\chi_{\parallel N}$ and $\chi_{\parallel S}$ denote the field-dependent spin susceptibility of the normal and superconducting states, respectively; $N_0$ is the density of states at the Fermi surface. For a BCS superconductor, electron spins are Heisenberg-like, i.e. no preferential direction. Under $H_\parallel$, the quasiparticle peaks are Zeeman split by $g\mu_B H_\parallel$ (ref. [10, 11]) ($g$ and $\mu_B$ are the electron g-factor and the Bohr magneton, respectively). The spin susceptibility of the superconducting state is zero ($\chi_{\parallel S} = 0$) since Cooper pairs are singlet [19, 20]. The spin susceptibility of the normal state is the Pauli paramagnetic susceptibility $\chi_{\parallel N} = N_0(g\mu_B)^2/2$. The transition occurs when $F_S = F_N$ at the familiar upper critical field $H_{c2}^\parallel = H_P$. It is first order since the derivative of the two free energies is discontinuous at $H_P$ (Fig. 1a) [10, 12].

On the other hand, in 2D NbSe$_2$ with Ising spins pinned by the strong intralayer SOC field [9, 13, 15], $H_{SO}(\gg H_\parallel)$, the in-plane field induces an electron spin magnetic moment of $\sim g\mu_B(\frac{H_\parallel}{H_{SO}})$ and a Zeeman splitting of $\sim g\mu_B H_\parallel \left(\frac{H_\parallel}{H_{SO}}\right) \ll g\mu_B H_\parallel$. This energy is ~ 0.1 meV even at $H_\parallel = 30$ T, which explains the negligible Zeeman splitting observed in the quasiparticle peaks. The Ising SOC also endows the in-plane spin susceptibility a van Vleck character. The van Vleck spin susceptibility is originated from virtual transitions involving states over an energy scale much larger than the superconducting gap $\Delta_0$ [22] (the band width, the spin-orbit splitting and $\Delta_0$ of NbSe$_2$ are on the order of 1 eV, 100 meV and 1 meV, respectively). As a result, the opening of a small superconducting gap at the Fermi surface has little effect on the spin susceptibility, i.e. $\chi_{\parallel S} \approx \chi_{\parallel N}$ (Ref. [20, 21]). The two free energies therefore touch smoothly to yield a continuous paramagnetic-limited transition at a much higher $H_{c2}^\parallel$ (Fig. 1a), as observed in our experiment.

Finally, we perform quantitative calculations of the superconducting gap as a function of in-plane magnetic field in trilayer NbSe$_2$ to validate the above physical picture. The results have been obtained by solving the self-consistent mean-field equation based on the *ab initio* band structure of trilayer NbSe$_2$ and on the assumption of an isotropic *s*-wave pairing potential [13, 15]. The latter assumption is consistent with studies of the superconducting gap of bulk NbSe$_2$ by the angle-resolved photoemission spectroscopy (ARPES) [33]. For comparison, we have also included results when the Ising SOC has been turned off numerically. Figure 3a shows the temperature dependence of the ratio of the susceptibilities $\chi_{\parallel S}/\chi_{\parallel N}$ in the low-field limit. Whereas $\chi_{\parallel S}/\chi_{\parallel N}$ goes to zero at low temperature in the absence of SOC, as expected for a BCS superconductor [19], $\chi_{\parallel S}/\chi_{\parallel N}$ remains close to unity even at low temperatures when SOC is included. Figure 3b depicts the field dependence of the gap. In the presence of SOC (main panel), the gap decreases to zero continuously with field at all temperatures, in good agreement with our experiment (Fig. 2c). In contrast, the transition becomes first order at low temperatures in the absence of SOC (inset, Fig. 3b), in agreement with the behavior of a BCS superconductor (Fig. 2e) [10-12].

In conclusion, we have observed an unusual continuous paramagnetic-limited superconductor-normal metal transition in 2D NbSe$_2$ at low temperature by tunneling spectroscopy. Our results have provided strong evidence for significant spin susceptibility of 2D NbSe$_2$ in the zero temperature limit, which is originated from the antisymmetric Ising SOC. The findings together with the developed junction devices have paved the path for probing exotic superconducting phenomena such as equal-spin Andreev reflection, proximity phenomena and topological superconductivity [13, 15, 16] in 2D NbSe$_2$.

**References**

1. Bauer, E. & Sigrist, M. Non-centrosymmetric Superconductors: Introduction and Overview (eds. Ernst Bauer & Sigrist, M.) (Springer-Verlag Berlin Heidelberg, 2012).
2. Bauer, E. et al. Heavy fermion superconductivity and magnetic order in noncentrosymmetric CePt3Si. *Physical Review Letters* **92**, 027003 (2004).
3. Yogi, M. et al. Evidence for novel pairing state in noncentrosymmetric superconductor CePt3Si: Si-29-NMR knight shift study. *Journal of the Physical Society of Japan* **75**, 013709 (2006).
4. Yip, S. Noncentrosymmetric Superconductors. *Annual Review of Condensed Matter Physics* **5**, 15-33 (2014).
5. Smidman, M., Salamon, M.B., Yuan, H.Q. & Agterberg, D.F. Superconductivity and spin–orbit coupling in non-centrosymmetric materials: a review. *Reports on Progress in Physics* **80**, 036501 (2017).
6. Cao, Y. et al. Quality Heterostructures from Two-Dimensional Crystals Unstable in Air by Their Assembly in Inert Atmosphere. *Nano Letters* **15**, 4914-4921 (2015).
7. Tsen, A.W. et al. Nature of the quantum metal in a two-dimensional crystalline superconductor. *Nat Phys* **12**, 208-212 (2016).
8. Ugeda, M.M. et al. Characterization of collective ground states in single-layer NbSe2. *Nat Phys* **12**, 92-97 (2016).
9. Xi, X. et al. Ising pairing in superconducting NbSe2 atomic layers. *Nat Phys* **12**, 139-143 (2016).
10. Meservey, R. & Tedrow, P.M. Spin-polarized electron-tunneling *Physics Reports-Review Section of Physics Letters* **238**, 173-243 (1994).
11. Meservey, R., Tedrow, P.M. & Bruno, R.C. Tunneling measurements on spin-paired superconductors with spin-orbit scattering. *Physical Review B* **11**, 4224-4235 (1975).
12. Adams, P.W., Herron, P. & Meletis, E.I. First-order spin-paramagnetic transition and tricritical point in ultrathin Be films. *Physical Review B* **58**, R2952-R2955 (1998).
13. Zhou, B.T., Yuan, N.F., Jiang, H.-L. & Law, K.T. Ising superconductivity and Majorana fermions in transition-metal dichalcogenides. *Physical Review B* **93**, 180501 (2016).
14. Yuan, N.F., Mak, K.F. & Law, K.T. Possible topological superconducting phases of MoS 2. *Physical review letters* **113**, 097001 (2014).
15. He, W.-Y., Zhou, B.T., He, J.J., Zhang, T. & Law, K. Nodal Topological Superconductivity in Monolayer NbSe2. *arXiv preprint arXiv:1604.02867* (2016).
16. Hsu, Y.-T., Vaezi, A., Fischer, M.H. & Kim, E.-A. Topological superconductivity in monolayer transition metal dichalcogenides. *Nature Communications* **8**, 14985 (2017).
17. Tinkham, M. Introduction to superconductivity (McGraw-Hill Book Co., New York, 2004).
18. Liu, C.X. Unconventional Superconductivity in Bilayer Transition Metal Dichalcogenides. *Physical Review Letters* **118**, 087001 (2017).
19. Frigeri, P.A., Agterberg, D.F. & Sigrist, M. Spin susceptibility in superconductors without inversion symmetry. *New Journal of Physics* **6**, 115 (2004).

20. Gor'kov, L.P. & Rashba, E.I. Superconducting 2D System with Lifted Spin Degeneracy: Mixed Singlet-Triplet State. *Physical Review Letters* **87**, 037004 (2001).
21. Yip, S.K. Two-dimensional superconductivity with strong spin-orbit interaction. *Physical Review B* **65**, 144508 (2002).
22. Wakatsuki, R. & Law, K.T. Proximity effect and Ising superconductivity in superconductor/transition metal dichalcogenide heterostructures. *arXiv preprint arXiv:1604.04898* (2016).
23. Lu, J.M. et al. Evidence for two-dimensional Ising superconductivity in gated MoS2. *Science* **350**, 1353 (2015).
24. Saito, Y. et al. Superconductivity protected by spin-valley locking in ion-gated MoS2. *Nature Physics* **12**, 144-149 (2015).
25. Nam, H. et al. Ultrathin two-dimensional superconductivity with strong spin–orbit coupling. *Proceedings of the National Academy of Sciences* **113**, 10513-10517 (2016).
26. Xi, X. et al. Strongly enhanced charge-density-wave order in monolayer NbSe2. *Nature nanotechnology* **10**, 765-769 (2015).
27. Daghero, D. & Gonnelli, R.S. Probing multiband superconductivity by point-contact spectroscopy. *Superconductor Science and Technology* **23**, 043001 (2010).
28. Webb, G.W., Marsiglio, F. & Hirsch, J.E. Superconductivity in the elements, alloys and simple compounds. *Physica C: Superconductivity and its Applications* **514**, 17-27 (2015).
29. Bianchi, A. et al. First-Order Superconducting Phase Transition in CeCoIn5. *Physical Review Letters* **89**, 137002 (2002).
30. Radovan, H.A. et al. Magnetic enhancement of superconductivity from electron spin domains. *Nature* **425**, 51 (2003).
31. Lortz, R. et al. Calorimetric Evidence for a Fulde-Ferrell-Larkin-Ovchinnikov Superconducting State in the Layered Organic Superconductor kappa-(BEDT-TTF)2Cu(NCS)2. *Physical Review Letters* **99**, 187002 (2007).
32. Zocco, D.A., Grube, K., Eilers, F., Wolf, T. & Löhneysen, H.v. Pauli-Limited Multiband Superconductivity in KFe2As2. *Physical Review Letters* **111**, 057007 (2013).
33. Kiss, T. et al. Charge-order-maximized momentum-dependent superconductivity. *Nature Physics* **3**, 720-725 (2007).

**Methods**

**Device fabrication**

The normal metal-insulator-superconductor (NIS) junctions employed in this experiment were fabricated using the dry transfer method [9]. Atomically thin flakes of NbSe$_2$ were mechanically exfoliated onto flexible polydimethylsiloxane (PDMS) substrates from bulk NbSe$_2$ single crystals, which were grown by the vapour transport method. The thickness of the NbSe$_2$ flakes was first estimated by their optical contrast on PDMS and then verified by their characteristic thickness dependent superconducting transition temperatures and/or Raman shear mode frequencies [26]. The identified flakes of appropriate thickness, size and shape were then transferred onto the Si substrates with pre-patterned metal electrodes. Two types of tunnel junction devices have been explored in this study. For the first type, split electrodes of Ti/Pt (4 nm/50 nm) were first deposited on Si/SiO$_2$ (300 nm) substrates by the standard photolithography and e-beam evaporation method. The substrates were then cleaned under an ozone environment, followed by e-beam evaporation of a layer of Al of ~ 0.3 – 1.5 nm thickness (see Supplementary Sect. 3.1 for optimization of the tunnel barrier thickness). The Al layer was allowed to fully oxidize to form AlO$_x$ under ambient conditions for a few hours. In order to minimize contribution from the electrode resistance to the voltage drop across the junctions, the NbSe$_2$ flakes were aligned and transferred such that one of the sample edges is very close to where the electrodes are split (Fig. 1c). Finally a thin layer of hexagonal boron nitride (hBN) was transferred on top of the devices as a capping layer. Typical residual resistivity ratio (RRR) of our devices is about 10. For the second type of devices, the exfoliated NbSe$_2$ flakes were directly transferred onto pre-patterned Ti/Au (5 nm/50 nm) electrodes. A thin layer (3 – 4 layer) of MoS$_2$ tunnel barrier was then transferred onto the NbSe$_2$ sample, followed by patterning a Ti/Au (5 nm/50 nm) tunnel electrode by e-beam lithography. The typical RRR (~ 5) is lower in this type of devices. (See Supplementary Sect. 3.2, 4 and 5.3 for results from this type of devices.)

**Transport measurement**
Both four-point resistance and differential conductance have been measured. In the differential conductance measurements, the junction was biased through one of the split electrodes with a small ac bias current, superimposed on a dc bias current (the amplitude of the ac voltage was kept below 50 $\mu$V and the dc voltage was varied from -5 mV to 5 mV). Both the ac and dc voltage drop at the junction were measured through the other electrode of the split pair using a lock-in amplifier and a voltage preamplifier, respectively (See Supplementary Sect. 1 for the measurement schematic). The differential conductance $G$ was obtained as the ratio of the ac current to the ac voltage. The dc voltage dependence of $G$ is the differential tunneling conductance spectrum reported in the main text. Initial screening of the devices up to 9 Tesla was performed using a Physical Property Measurement System (PPMS). Measurements under high magnetic fields were performed using the 45 T hybrid magnet at the National High Magnetic Field Laboratory (NHMFL) in a $^3$He refrigerator with a $^3$He exchange gas. The hybrid magnet only allows the application of magnetic field between 11 T and 45 T for most of the temperatures in this study. To apply an in-plane magnetic field with a misalignment < 0.1 degrees, we aligned the devices on a rotating probe such that the four-point resistance was minimized when the magnetic field was fixed slightly below the in-plane upper critical field at the base temperature. This procedure utilizes the large anisotropy in the in-plane and out-of-plane upper critical fields. (Details of the alignment procedure are given in Supplementary Sect. 2.)

**Estimate of the in-plane orbital-limited upper critical field**

The in-plane orbital-limited upper critical field can be estimated following $H_{c2}^{\parallel}(T=0) = \frac{\sqrt{3}}{\pi}\frac{\Phi_0}{\xi_{\parallel} t}$ (ref. [17, 25]), where $\Phi_0$, $\xi_{\parallel}$ and $t$ are the flux quantum, in-plane coherence length and sample thickness, respectively. We obtained $H_{c2}^{\parallel} \approx 85$ T and 170 T, respectively, for trilayer and bilayer NbSe$_2$. In this calculation, we used $\xi_{\parallel} \approx 10$ nm [7] and $t \approx 1.3$ nm [9, 26] for trilayer NbSe$_2$, and $t \approx 0.65$ nm for bilayer NbSe$_2$. These fields far exceed the experimentally measured in-plane upper critical fields (~ 37 T) for both bilayer and trilayer samples. The superconductor-normal metal transition considered here is thus safely in the paramagnetic-limited regime. For simplicity we have neglected the orbital effects in our discussions.

**References for Methods**

7. Tsen, A.W. et al. Nature of the quantum metal in a two-dimensional crystalline superconductor. *Nat Phys* **12**, 208-212 (2016).
9. Xi, X. et al. Ising pairing in superconducting NbSe2 atomic layers. *Nat Phys* **12**, 139-143 (2016).
17. Tinkham, M. Introduction to superconductivity (McGraw-Hill Book Co., New York, 2004).
25. Nam, H. et al. Ultrathin two-dimensional superconductivity with strong spin–orbit coupling. *Proceedings of the National Academy of Sciences* **113**, 10513-10517 (2016).
26. Xi, X. et al. Strongly enhanced charge-density-wave order in monolayer NbSe2. *Nature nanotechnology* **10**, 765-769 (2015).

**Acknowledgments**

This research was supported by the ARO Award W911NF-17-1-0605 for sample and device fabrication and the US Department of Energy, Office of Basic Energy Sciences contract No. DESC0013883 for tunneling spectroscopy measurements. A portion of this work was performed at the National High Magnetic Field Laboratory, which is supported by National Science Foundation Cooperative Agreement No. DMR-1644779 and the State of Florida. The work in Hong Kong was supported by the Croucher Foundation, the Dr. Tai-chin Lo Foundation and the Hong Kong Research Grants Council through HKUST3/CRF/13G, C6026-16W and 16324216. The work in Lausanne was supported by the Swiss National Science Foundation. We also acknowledge support from the National Science Foundation under Award No. DMR-1645901 (E.S.), DMR-1420451 (K.K.), DMR-1410407 (Z.W.) and a David and Lucille Packard Fellowship and a Sloan Fellowship (K.F.M.).

**Figures and tables**

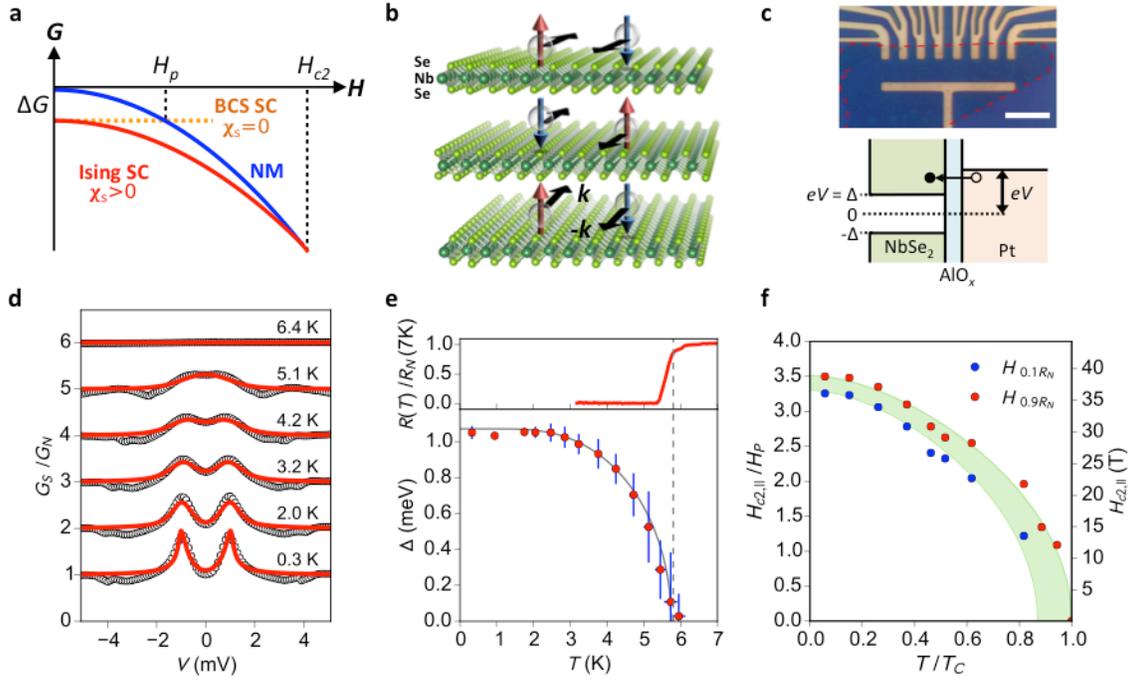

**Figure 1 | 2D NbSe₂ superconductor tunnel junction. a,** Magnetic-field dependence of the free energy of a normal metal (NM, blue line), BCS superconductor (SC) (orange dashed line) and an Ising SC (red line). $\Delta G$ is the condensation energy. The paramagnetic-limited transition is first order at $H_P$ if spin susceptibility $\chi_s = 0$ (BCS SC), and second order at a higher upper critical field $H_{c2}$ if spin susceptibility is significant $\chi_s > 0$ (Ising SC). **b,** Schematic of Ising spins in superconducting trilayer NbSe₂. In each monolayer, electrons of the K and K' pocket possess opposite momentum (in-plane arrows) and opposite spin (out-of-plane arrows). Electrons of the same momentum in adjacent weakly coupled layers have opposite spins since the two layers are inverted. **c,** Optical image of a typical NbSe₂ tunnel junction (top). It consists of Pt electrodes (yellow), a thin layer of AlO$_x$, NbSe₂ superconductor (outlined by red dashed lines), and hBN capping layer (blue). The scale bar corresponds to 10 μm. The energy diagram of the junction (bottom) shows that the tunneling current as a function of bias voltage $V$ measures the superconducting gap 2Δ. **d,** Differential tunneling conductance spectra (symbols) and comparison to the BTK model (red lines) of trilayer NbSe₂ at varying temperatures under zero magnetic field. **e,** Temperature dependence of the normalized four-point resistance (solid red line, upper panel) and the temperature dependence of the superconducting gap Δ from tunneling measurements (symbols, lower panel). The vertical and horizontal error bars of Δ correspond to the BTK fitting error and sample temperature fluctuations, respectively. The dashed line indicates the critical temperature $T_C$. The solid gray line is the prediction of the BCS theory with zero temperature gap $2\Delta_0 \approx 4.3 k_B T_C$. **f,** Temperature dependence of the critical field $H_{c2}^{\parallel}$ (red symbols) and of the superconducting transition width (shaded region between 90% (red) and 10% (blue) of the normal-state resistance). The critical field and the sample

temperature are normalized by the spin paramagnetic limit of the BCS theory $H_P = 10.8$ T and the critical temperature $T_C \approx 5.8$ K, respectively.

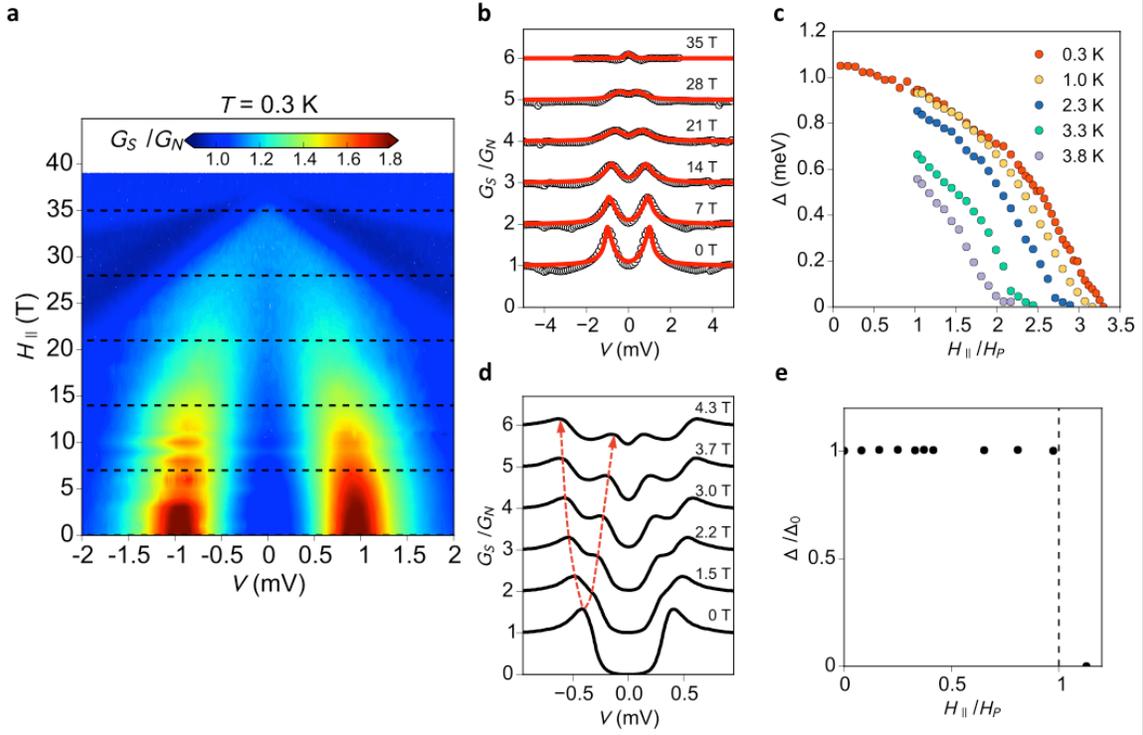

**Figure 2 | Tunneling spectroscopy under in-plane magnetic fields. a,** Contour plot of the differential conductance of trilayer NbSe$_2$ as a function of bias voltage ($V$, bottom axis) and in-plane field ($H_\parallel$, left axis) at 0.3 K. A continuous closing of the superconducting gap is visible when the field is increased. **b,** Differential tunneling conductance spectra (symbols) at selected fields (dashed lines in **a**) and comparison to the BTK model (red solid lines). The spectra are vertically displaced for clarity. **c,** The extracted superconducting gap from the tunneling spectra as a function of field at differing temperatures. **d, e,** Tunneling spectra at representative in-plane fields, vertically displaced for clarity (**d**), and field dependence of the superconducting gap (**e**) for aluminum thin films at low temperature are included for comparison. The dashed red lines in **d** illustrate the Zeeman splitting of the quasiparticle peak. Adapted from ref. 11, APS (**d**).

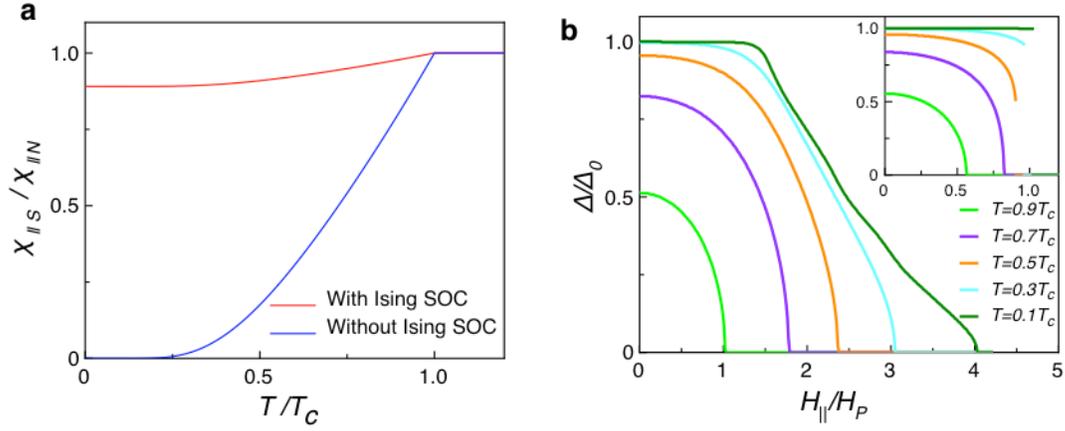

**Figure 3 | Calculations of the superconducting gap under in-plane magnetic fields. a,** In-plane spin susceptibility of trilayer NbSe$_2$ normalized by the normal-state value ($\chi_{\|S}/\chi_{\|N}$) in the low-field limit as a function of temperature with (red line) and without SOC (blue line). **b,** Superconducting gap normalized by the zero-temperature, zero-field value ($\Delta/\Delta_0$) as a function of in-plane field at differing temperatures. The gap closes continuously with increasing field at all temperatures. Inset shows the result without including SOC in the band structure of NbSe$_2$. The band gap closes abruptly with increasing field at low temperature (0.1 $T_C$).

**Table 1. Experimentally observed paramagnetic-limited superconductor-normal metal transitions.**

| | Al [10,11] | Be [12]* | V-Ti [10] | CeCoIn$_5$ [29,30]†‡ | κ-(BEDT-TTF)$_2$Cu(NCS)$_2$ [31]† | KFe$_2$As$_2$ [32]§¶ | 2D NbSe$_2$ |
|---|---|---|---|---|---|---|---|
| | (metal thin film) | (metal thin film) | (metal thin film) | (Heavy fermion SC) | (Organic SC) | (Pnictide SC) | |
| $H_{c2}/H_p$ | ≈ 1.0 | ≈ 1.1 | ≈ 1.1 | ≈ 0.9 | ≈ 1.0 | ≈ 0.6 | ≈ 3.5 |
| Zeeman splitting | Y | - | Y | - | - | - | N |
| Phase transition order ($T \approx 0$) | 1st | 1st | 1st | 1st | 1st | 1st | 2nd |

Experimental methods in addition to tunneling measurements: *Transport; †Specific heat; ‡Magnetization; §Thermal expansion; ¶Magnetostriction.